\documentclass[journal]{IEEEtran}
\usepackage{cite}
\usepackage{multirow}
\usepackage{upgreek}
\usepackage{placeins}
\usepackage{nicefrac}

\ifCLASSINFOpdf
  \usepackage[pdftex]{graphicx}
 \graphicspath{{Figures/}}
  \DeclareGraphicsExtensions{.pdf,.jpg,.png}
\fi

\usepackage{amsmath}

\ifCLASSOPTIONcompsoc
  \usepackage[caption=false,font=normalsize,labelfont=sf,textfont=sf]{subfig}
\else
  \usepackage[caption=false,font=footnotesize]{subfig}
\fi

\usepackage{stfloats}

\usepackage{url}
\usepackage{multicol}

\usepackage{xcolor}

\begin{document}

\title{Printed tapered leaky-wave antennas for W-band frequencies}

\author{Andreas~E.~Olk, 
		Mingkai~Liu, 
        and David~A.~Powell,~\IEEEmembership{Senior Member,~IEEE}
\thanks{A. E. Olk  is affiliated with the University of New South Wales, Canberra, Australia and with IEE S.A., Luxembourg.}
\thanks{Mingkai Liu is affiliated with the Australian National University, Canberra, Australia.}
\thanks{D. A. Powell is affiliated with the University of New South Wales, Canberra, Australia.}
\thanks{e-mail of corresponding author A. E. Olk: andreas.olk@iee.lu }

}


\maketitle

\begin{abstract}

Despite their great potential in communication and sensing applications, printed leaky-wave antennas have rarely been reported at mm-wave frequencies. In this paper, tapered leaky-wave antennas operating at 80\,GHz are designed, fabricated and experimentally characterized. While many continuous leaky-wave antennas use subwavelength strips or other comparably small elements, in this work, the surface impedance is discretized very coarsely using only three square patches per period. With this architecture, a wide range of surface reactance can be achieved while maintaining a minimum feature size of the metallic pattern that is feasible for printed circuit fabrication. 
Another advantage of this approach over existing works in the mm-wave frequency range is that it allows precise engineering of the aperture illumination. We demonstrate this by applying amplitude tapering for side lobe suppression. A comprehensive experimental study is presented, including near-field and far-field measurements. Therewith, we verify the designed aperture illumination and we reveal the origin of spurious far-field features. Side lobes are effectively suppressed and spurious radiation is reduced to ${-18}$\,dB compared to the main lobe.
 
\end{abstract}


%
\IEEEpeerreviewmaketitle

\section{Introduction}
\label{sec:LWAintroduction}

\IEEEPARstart{F}{or} several decades, leaky-wave antennas have attracted significant attention as they offer unique properties such as low-profile, high directivity and frequency scanning capabilities \cite{guhaMicrostripPrintedAntennas2011,jacksonLeakywaveAntennas2008}. 
One of the simplest forms of leaky-wave antennas are 1D sinusoidally-modulated reactance surfaces. They are based on the initial solution of A. Oliner \cite{olinerGuidedWavesSinusoidallymodulated1959} and have been studied intensively \cite{patelPrintedLeakyWaveAntenna2011,yangTaperedUnitCell2018,panaretosLeakyWaveAntennasBased2016,oraiziDesignWidebandLeakyWave2018,kuznetcovPrintedLeakyWaveAntenna2019}. By varying the modulation depth, the leakage rate can be controlled, while the propagation constant does not vary significantly. Therefore, sinusoidally-modulated leaky-wave antennas can be designed for a wide range of aperture size and beamwidth. Additionally, tapering techniques can be applied to control the side lobe level \cite{ranceGeneralizedArrayFactor2015,yangTaperedUnitCell2018,gomez-torneroFFTSynthesisRadiation2010}. Due to the open-stopband, they are suited mainly for moderate off-broadside angles \cite{patelPrintedLeakyWaveAntenna2011,baccarelliNewBrillouinDispersion2007}.


For novel applications in wireless communication and radar sensing, high frequencies of operation in the millimeter wave (mm-wave) band are increasingly used. Although leaky-wave antennas have shown great potential for such applications, very limited work has been reported in the W-band (75-110\,GHz) or above. Fabrication tolerances \cite{fischer_causes_2013,olkHighEfficiencyRefractingMillimeterWave2020} as well as material losses \cite{fischer_causes_2013} and conductor roughness \cite{goldPhysicalSurfaceRoughness2017,gopalakrishnanStudyEffectSurface2016,olkHighEfficiencyRefractingMillimeterWave2020} are more problematic at these frequencies. The design of efficient mm-wave devices is therefore challenging. Recently, several mm-wave leaky-wave antenna concepts have been proposed, but they mostly include complex non-planar manufacturing processes. In Ref. \cite{baiSinusoidallyModulatedLeakyWave2016} for instance, a
leaky-wave antenna based on an open corrugated waveguide filled with dielectric is demonstrated. Additionally, the authors of Ref. \cite{chengMillimeterWaveLowTemperature2014} propose a leaky-wave antenna made from low-temperature co-fired ceramics (LTCC) which is a costly high performance material. Other works include leaky-wave antennas based on perforated dielectrics \cite{mondalLeakyWaveAntennaUsing2016} or complex substrate-superstrate configurations \cite{gomez-torneroApplicationHighgainSubstratesuperstrate2005}. Other challenges in the design of mm-wave leaky-wave antennas include the experimental characterization, as well as the design of an efficient surface wave launcher \cite{maCollimatedSurfaceWaveExcitedHighImpedance2017,oraiziDesignWidebandLeakyWave2018,kuznetcovPrintedLeakyWaveAntenna2019}.

Most continuous leaky-wave antennas use subwavelength strips \cite{patelPrintedLeakyWaveAntenna2011,yangTaperedUnitCell2018,panaretosLeakyWaveAntennasBased2016} or other comparably small elements \cite{oraiziDesignWidebandLeakyWave2018}, typically with 10 or more elements per modulation period. This can lead to small feature sizes and a large number of elements for a given aperture. 
Such antenna architectures with deeply subwavelength structures are therefore impractical to fabricate for operation in the millimeter wave range. Other periodic leaky wave antennas, particularly those for open-stopband suppression, such as periodically loaded transmission lines \cite{hammesPeriodicHighSensitive2019,paulottoNovelTechniqueOpenStopband2009} or slotted closed waveguides \cite{liuSimpleTechniqueOpenStopband2018,maCollimatedSurfaceWaveExcitedHighImpedance2017} use only one or two (potentially connected) elements per period. This results in structures which are geometrically simple and potentially easy to scale to higher frequencies. However, many of these periodic leaky-wave antennas  are based on high permittivity substrates \cite{kuznetcovPrintedLeakyWaveAntenna2019,liuSimpleTechniqueOpenStopband2018,maCollimatedSurfaceWaveExcitedHighImpedance2017} which are disadvantageous for mm-wave due to the high dielectric losses. 
Goubau line leaky-wave antennas feature relatively coarse connected structures, low losses and they were proposed recently for mm-wave applications  \cite{rudramuniGoubauLineLeakyWaveAntenna2018,tangContinuousBeamSteering2017}. They were experimentally investigated at around 40\,GHz \cite{sanchez-escuderosPeriodicLeakyWaveAntenna2013}. As they are made without a ground plane, they  feature omnidirectional radiation pattern \cite{tangContinuousBeamSteering2017,rudramuniGoubauLineLeakyWaveAntenna2018} or they have to be combined with a parallel oriented reflecting plane \cite{sanchez-escuderosPeriodicLeakyWaveAntenna2013}.

In this paper, a modulated reactance antenna is proposed for W-band frequencies. The continuous surface reactance is discretized very coarsely using three square patches per period. We show that this structure is particularly suited for higher frequencies in the mm-wave range as it enables a wide range of reactance modulation while maintaining a minimum feature size that is compatible with printed circuit fabrication. Due to the fact that the band diagram deviates significantly from the analytical solution, an analysis of the unit-cell with eigenmode solvers is performed. Similar to conventional sinusoidally-modulated reactance antennas with more subwavelength discretization, the leakage rate can be controlled by varying the modulation depth. To demonstrate this capability, we design several different tapered leaky-wave antennas with aperture illumination according to the Taylor one-parameter distribution \cite{maillouxPhasedArrayAntenna2005}. Using this illumination distribution, the trade-off between side lobe level and main lobe width can be controlled effectively. For moderate off-broadside angles and for angles close to stopband, the fidelity of the radiation pattern is very high and side lobes are suppressed effectively. The experimental demonstration includes near-field and far-field characterization with which we verify the designed aperture distribution and show the origin of spurious far-field features.

\section{Continuous leaky-wave antenna with coarse discretization}
\label{sec:coarseSMRS}

To design a sinusoidally-modulated reactance antenna, a surface impedance of the form
\begin{equation}
    \eta_{surf}(z)=j\eta_0 X' \left[1+M(z) \cos \left ( \frac{2\pi z}{d_p}+\phi_{0} \right) \right]
\end{equation}
is required \cite{patelPrintedLeakyWaveAntenna2011}. Here, $\eta_0$ is the free space impedance, $X'$ is the normalized average surface reactance, $M(z)$ the modulation depth and $d_p$ the period of modulation. In this section and in Section \ref{sec:uniform_modulation}, we consider the case of uniform modulation, i.e. $M(z)$ is constant. A general and detailed design recipe for sinusoidally-modulated reactance surfaces can be found in Ref. \cite{patelPrintedLeakyWaveAntenna2011}. This design procedure is widely used throughout the literature \cite{yangTaperedUnitCell2018,panaretosLeakyWaveAntennasBased2016,oraiziDesignWidebandLeakyWave2018}. One of the main assumptions in the procedure is that the surface reactance is quasi continuous, i.e.~it is discretized with subwavelength elements. In many previous works, the reactance surface consists of an array of conductor strips placed on a grounded dielectric substrate \cite{yangTaperedUnitCell2018,panaretosLeakyWaveAntennasBased2016,oraiziDesignWidebandLeakyWave2018}. One modulation period usually consists of at least 10 strips and the relative permittivity of the substrate is $\epsilon_r=6.15$ or larger \cite{yangTaperedUnitCell2018,panaretosLeakyWaveAntennasBased2016,oraiziDesignWidebandLeakyWave2018}.

\begin{figure}[htbp]
	\centering
	\includegraphics[width=0.42\textwidth]{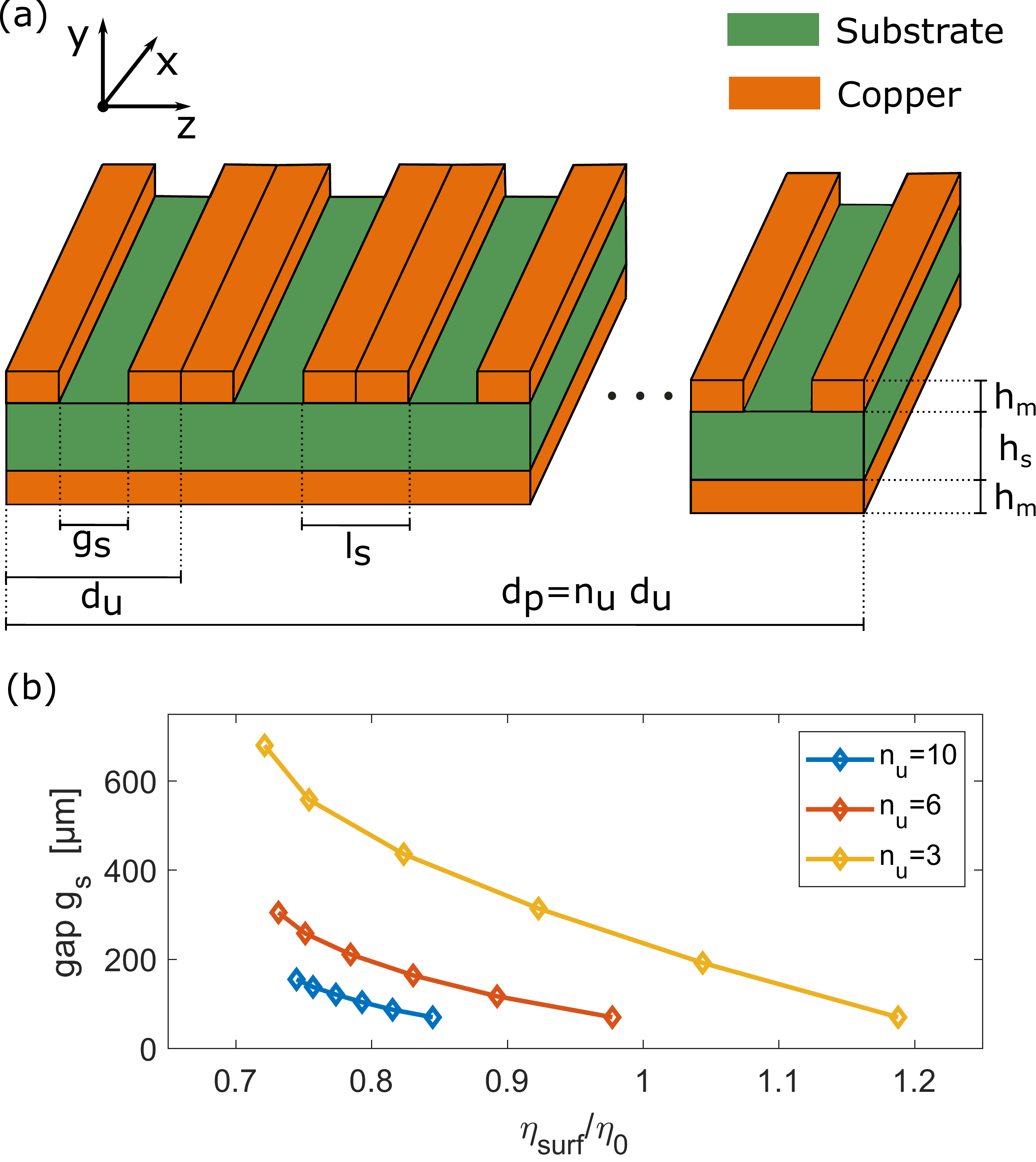}
	\caption{(a) Conventional architecture of sinusoidally modulated reactance surfaces \cite{patelPrintedLeakyWaveAntenna2011,olinerLeakywaveAntennas1993} and (b) the corresponding lookup table using $\epsilon_r=3.0$ and a frequency of 80\,GHz. The fixed geometrical parameters are $d_p=2.25$\,mm, $h_m=35\,\upmu$m and $h_s=508\,\upmu$m.}
	\label{fig:slottedSMRS}
\end{figure}

This architecture with subwavelength metallic elements is impractical for higher frequency applications as it can lead to feature sizes which are too small for printed circuit fabrication and the dielectric losses of substrates with high permittivity are significant. Here we give an example, showing that coarse discretization can ameliorate these issues. We consider the parameterized geometry shown in Figure~\ref{fig:slottedSMRS}~(a) with a low loss substrate of $\epsilon_r=3.0$ and $h_s=508\,\upmu$m and a frequency of operation of 80\,GHz. A rectangular piece of the reactance surface of width $d_u$ that contains one slit is referred to as one unit cell. The $n_u$ unit cells which correspond to one period $d_p$ of the modulation constitute a supercell. At mm-wave frequencies, losses introduced by the roughness of copper traces have a significant effect on the effective permittivity of transmission lines \cite{goldPhysicalSurfaceRoughness2017,olkHighEfficiencyRefractingMillimeterWave2020,curranMethodologyCombinedModeling2010}. In our design, we assume that the roughness of copper traces is 1.5\,$\upmu$m. According to the model reported in \cite{goldPhysicalSurfaceRoughness2017}, this means that the effective conductivity at 80\,GHz is $2\times 10^6\,$S/m. 
 
The angle of radiation can be approximated with
\begin{equation}
    \sin(\theta_{n})= k_{z,n}/k_0 \approx \sqrt{1+X'^2} + \frac{2 \pi n}{k_0 d_p}.
\end{equation}
where $k_0$ is the free space wave number and $k_{z,n}$ is the wave number of the $n$th spatial harmonic \cite{patelPrintedLeakyWaveAntenna2011}. Sinusoidally modulated reactance surfaces typically radiate with the $n=-1$ spatial harmonic. The average reactance $X'$ is set to 1.2. It was found from full-wave simulation (CST Microwave Studio \cite{cst}) that using this value minimizes spurious scattering at the junction of the feed and the sinusoidally-modulated reactance surface. This is in line with other works which used the same value \cite{patelPrintedLeakyWaveAntenna2011,yangTaperedUnitCell2018,panaretosLeakyWaveAntennasBased2016}. More detailed comments can be found in Section \ref{sec:contacts}. We aim to design antennas for moderate radiation angles and we choose $\theta_{n=-1}=-11^\circ$. Therefore, the period of modulation is set to 2.25\,mm. Using a commercial eigenmode solver ({Comsol Multiphysics} \cite{comsol}), the surface impedance of a rectangular patch in an infinite array of identical neighbors can be determined with the relation
\begin{equation}
    \eta_{unit}=-j\eta_0\sqrt{\left|\frac{k_z}{k_0}\right|^2-1}
\end{equation}
from Ref.~\cite{patelPrintedLeakyWaveAntenna2011}. Here, the $z$-direction is assumed to be the direction of propagation along the reactance surface and $k_z$ is the z-component of the wave number of an unmodulated reactance surface. We limit the minimum feature size of the design to 70\,$\upmu$m, which is a common requirement of commercial circuit board manufacturers. It is determined by the etching accuracy which is on the order of half the copper thickness, typically 18\,$\upmu$m. Due to this limitation, the gap between copper traces $g_s$ and the line width $l_s$ are constrained by $g_s\geq 70\,\upmu$m and $l_s=d_u-g_s\geq 70\,\upmu$m, respectively. Therefore, for different levels of discretization $n_u$, different minimum and maximum gap sizes apply. These limiting values for $n_u=10,6$ and $3$ are shown in Table \ref{tab:limits}. The resulting lookup table that relates the normalized surface reactance $\eta_{surf}/\eta_0$ to the gap width is shown in Figure~\ref{fig:slottedSMRS}~(b). It is discernible that the feasible reactance range is very small for $n_u=10$ and substantially improves for coarser discretization. 

\begin{table}
\begin{center}
\caption{Parameters for the array of strips}
\label{tab:limits}
\begin{tabular}{ |c | c | c | c|}
\hline
 $n_u$ & $d_u \, [\upmu \text{m}]$  & $g_{s,min} \, [\upmu \text{m}]$ & $g_{s,max} [\upmu \text{m}]$ \\ \hline \hline
 10 & 225 & 70 & 155 \\  
 6 & 375 & 70 & 305 \\ 
 3 & 750 & 70 & 680 \\ \hline
\end{tabular}
\end{center}
\end{table}

\begin{figure}[htbp]
	\centering
	\includegraphics[width=0.42\textwidth]{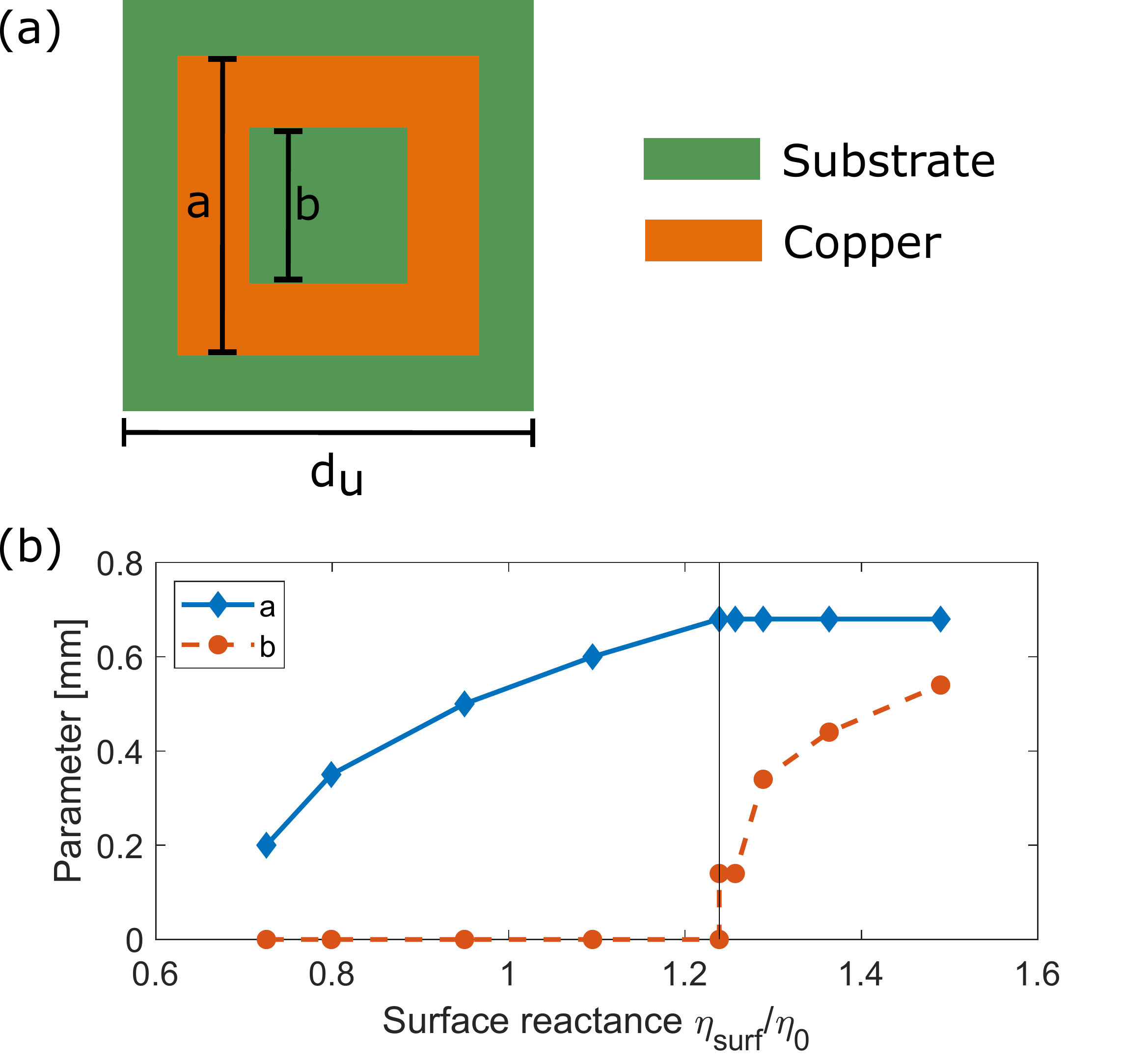}
	\caption{Proposed unit cell geometry with (a) top view and (b) lookup table for operation at 80\,GHz. The fixed geometrical parameters are $d_u=0.75$\,mm, $h_m=35\,\upmu$m and $h_s=508\,\upmu$m.}
	\label{fig:unit_patch}
\end{figure}

Given that an average reactance of $X'=1.2$ is required for efficiently feeding the surface wave onto the modulated reactance array, the impedance range of the coarsely discretized  metal strips is still too small. Therefore, we use an alternative geometry composed of isotropic square patches with side length $a$. Square patches are often used for 2D surface wave antennas with high directivity \cite{casalettiPolarizedBeamsUsing2016,minattiModulatedMetasurfaceAntennas2015,fongScalarTensorHolographic2010,pandiDesignScalarImpedance2015}, for instance in space applications \cite{casalettiPolarizedBeamsUsing2016,minattiModulatedMetasurfaceAntennas2015}. We extend the reactance range of the square patches to higher values by cutting an inner square with side length $b$ out of some of the patches. Increasing the cut out region results in higher inductance and thus higher surface reactance as it creates narrow lines. This geometry is shown in Figure~\ref{fig:unit_patch}~(a). We determine a lookup table that relates $a$ and $b$ to the surface reactance of the cell and we limit the minimum feature size to 70\,$\upmu$m as previously, i.e. $a\leq 680\,\upmu$m and $b\leq 540\,\upmu$m. The result is shown in Figure~\ref{fig:unit_patch}~(b) and shows a feasible reactance range from 0.73\,$\eta_0$ to 1.49\,$\eta_0$, which is larger than the reactance range achievable with strips having the same minimum feature size.

\section{Analysis of uniformly modulated reactance surfaces}
\label{sec:uniform_modulation}

\begin{figure}[htbp]
	\centering
	\includegraphics[width=0.42\textwidth]{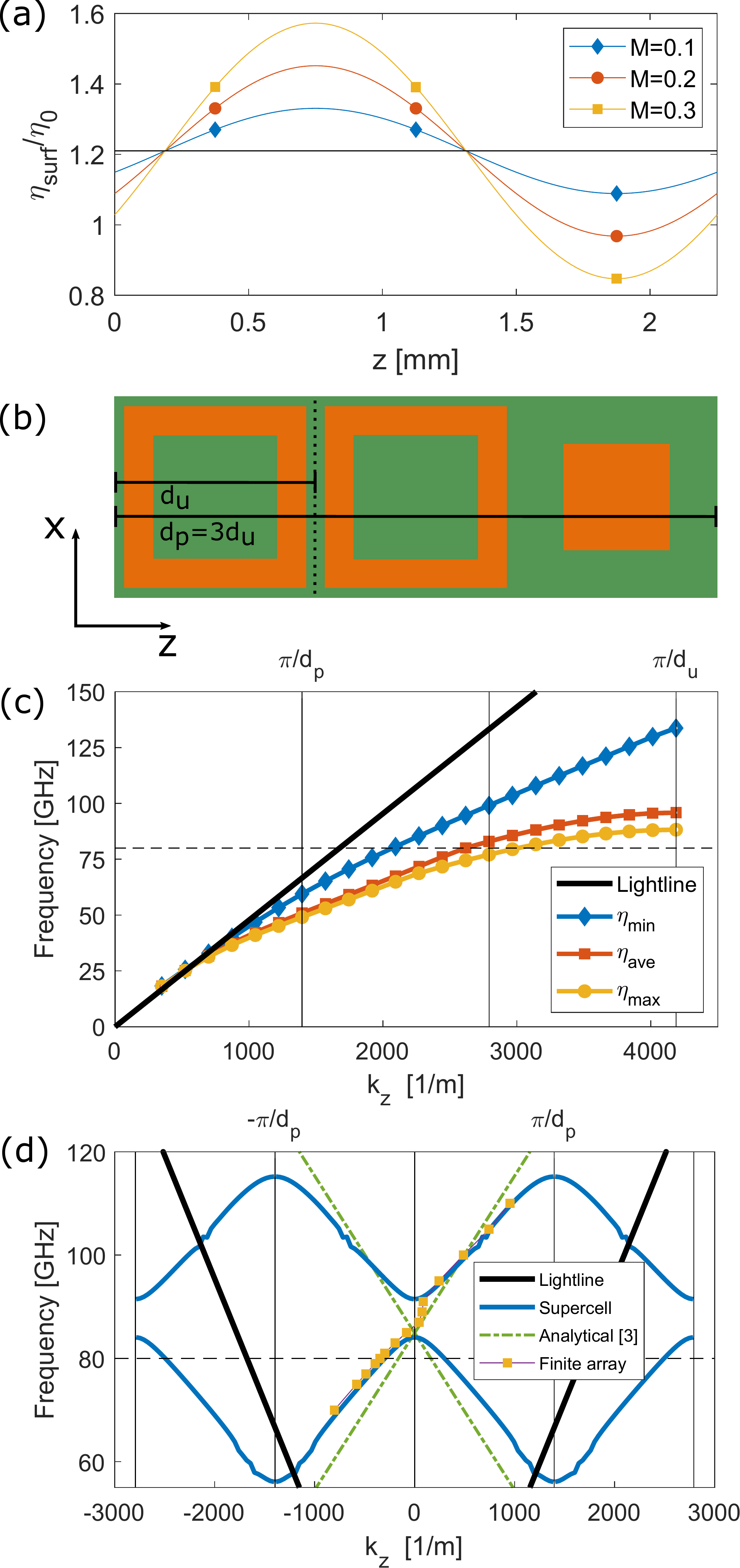}
	\caption{Supercell composed of three unit cells: (a) sinusoidal reactance modulation, (b) geometry, (c) dispersion of individual unit cells. (d) Band diagram of the supercell (solid blue line) in comparison with the analytical solution \cite{olinerGuidedWavesSinusoidallymodulated1959} (dashed green line) and the simulation result of the finite antenna array (yellow markers). The design frequency is marked with a horizontal dashed line.}
	\label{fig:supercell_and_band}
\end{figure}

Three metallic patches form one period of the sinusoidally-modulated reactance surface. The modulation $M$ is applied along the $z$ direction. In Figure \ref{fig:supercell_and_band}~(a), the ideal continuous surface reactance is shown as a solid curve together with the discretized values as markers for different values of $M$. The phase offset $\phi_0$ is set to ${-30^\circ}$, such that the sampled impedance values avoid the maximum peak, increasing the maximum modulation strength for an achievable range of impedance values. We note that this capability of reducing the required reactance with a suitable choice of $\phi_0$ is particular to coarsely discretized leaky wave antennas. The supercell layout for the reactance surface with maximum achievable modulation $M=0.3$ is shown in Figure~\ref{fig:supercell_and_band}~(b). The dispersion curve for the single patch with the maximum, minimum and average surface reactance are shown in Figure~\ref{fig:supercell_and_band}~(c). Due to the period of the homogeneous single patch array $d_u$, the corresponding dispersion curve runs from 0 to $k_z=\pi/d_u=4205\,m^{-1}$ which is marked by a vertical black line. All dispersion curves for a single patch array are entirely below the lightline (black curve). This confirms that without modulation, such an array is not radiating. The design frequency 80\,GHz is marked with a horizontal dashed line.

Similar to the analysis of the single patch array, the supercell is analyzed with a commercial full-wave eigenmode solver using periodic boundary conditions. From this simulation, the band diagram which is shown as blue curve in Figure~\ref{fig:supercell_and_band}~(d) is extracted. The supercell with its period of $d_p =3d_u$ folds back at multiples of $\pi/d_p=\pi/(3d_u)$. From this band diagram, it is discernible, that around the design frequency which is marked by a black horizontal dashed line, the supercell array solely radiates a backward wave with the $n=-1$ mode. The radiation of the $n=-1$ harmonic is used for all leaky-wave antenna designs in this work. Parasitic radiation from other spatial harmonics is expected to be effectively suppressed. An open-stopband occurs between 84.5 and 91.5\,GHz. An analytical approximation for the bandstructure of a periodic reactance surface (Eqs.~17 and 18 in Ref.~\cite{olinerGuidedWavesSinusoidallymodulated1959}) is plotted with a dashed green line. This analytical solution deviates strongly, even far away from the open-stopband. This confirms our assertion that for the accurate design of coarsely discretized modulated reactance surfaces, numerical design techniques are indispensable.

\begin{figure}[htbp]
	\centering
	\includegraphics[width=0.42\textwidth]{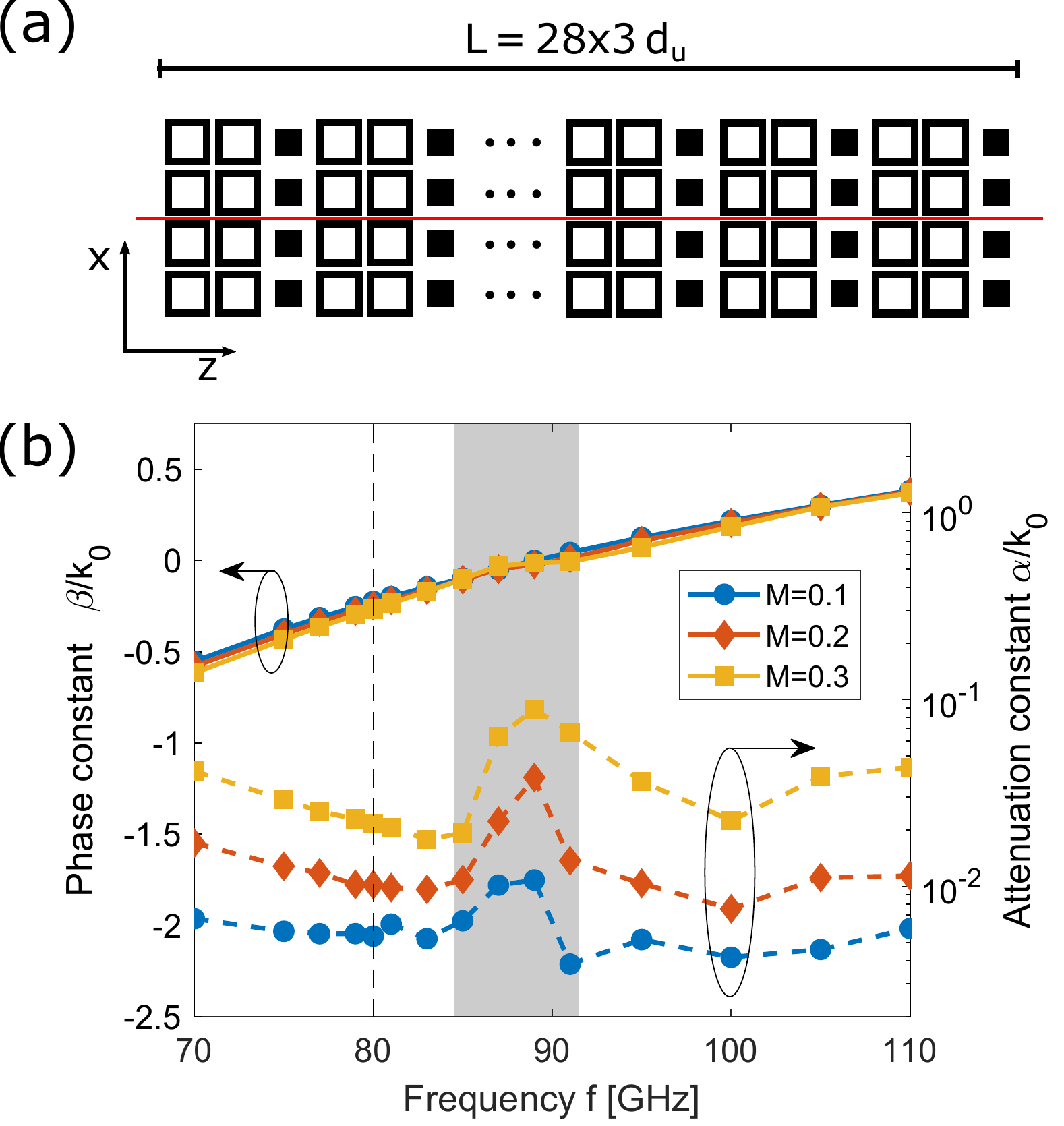}
	\caption{(a) leaky-wave antenna with constant reactance modulation $M=0.3$ and (b) phase constant $\beta$ and attenuation $\alpha$ evaluated for different modulation depths $M$. The grey shaded area indicates the stopband and the design frequency is marked with a vertical dashed line.}
	\label{fig:prop_const}
\end{figure}

The antennas designed in this work are composed of 4 identical patches in the $x$-direction and 28$\times$3 patches with modulated impedance along the $z$-direction (Figure~\ref{fig:prop_const}~(a)). In this configuration, the first and fourth patch of each row in the $x$-direction have fewer neighbors and therefore have a slightly different surface impedance from the infinite structure. After choosing a suitable supercell configuration based on the full-wave eigenmode solver, we evaluate the complex propagation constant $k_{z,n=-1}=\beta+i\alpha$ of the finite array using a commercial frequency domain solver (CST Microwave Studio \cite{cst}). Here, $\beta$ indicates the phase constant and $\alpha$ indicates the attenuation constant of the surface wave. The attenuation $\alpha$ contains contributions caused by leakage $\alpha_l$, dissipation $\alpha_d$ and reflections $\alpha_r$. Although significant losses occur here, we consider that leakage is the dominant effect around the design frequency. We show in Section~\ref{sec:LWA_experiments} that this approximation leads to accurate prediction of the normalized radiation pattern. 
In particular, to determine $\alpha$ and $\beta$, the transverse component of the magnetic field $H_y$ along a line in the middle of the array (marked in red in Figure~\ref{fig:prop_const}~(a)) and at a depth of half the substrate thickness $y=-h_s/2$ was fitted with the equation $\exp({j(\beta + j\alpha)z)}$. More details are given in Appendix~\ref{sec:appendix_1}. We note that alternatively the generalized-pencil-of-function approach \cite{huaMatrixPencilMethod1990} can be used to determine the propagation constant. This approach is insensitive to noise and it was used for the synthesis of linear arrays \cite{liuSynthesisNonuniformlySpaced2015} and more recently in periodic leaky-wave antennas \cite{comiteDirective2DBeam2021}. It can also be very accurate for estimating the attenuation within the stopband, which is not required for the synthesis of the antennas presented here.

The resulting phase constant $\beta$ and attenuation constant $\alpha$ are shown in Figure~\ref{fig:prop_const}~(b). The phase constant $\beta$ gradually increases for higher frequencies, but it remains relatively constant with variation of the modulation depth from $M=0.1$ to $M=0.3$. Below the stopband, marked by the grey shaded area, the antenna radiates backward waves and above the stopband forward waves. At the design frequency which is marked with a vertical dashed line, the attenuation constant $\alpha$ can be adjusted between 0 and $0.022\,k_0$.  Consequently, the independent control of propagation constant and leakage rate, which is an advantage of sinusoidally-modulated reactance antennas, is maintained also in this particular case of coarse discretization. We will make use of this property in the next section by applying tapering. The leakage rate is also dependent on frequency, particularly for strong modulation. This is one of the limiting factors for the bandwidth of the antenna. We also note that the choice of the design frequency is quite arbitrary. The presented unit cell architecture can also used above the stopband, e.g. around 100\,GHz where the attenuation is only slightly different. More details are given in Section \ref{sec:s_parameters} and Section \ref{sec:leaky_wave_farfield}.  In order to compare the propagation constant of the finite size array with the supercell simulation, the result for M=0.3 (yellow rectangular markers) is also plotted into Figure~\ref{fig:supercell_and_band}~(d). Outside the stopband, the discrepancy between the two methods is relatively small and is further reduced if more than four identical patches are aligned along the $x$ direction.  Using the finite size array, both $\alpha$ and $\beta$ are also obtained within the stopband and a strong increase of $\alpha$ is discernible which is due to reflections. Using 4 identical patches in $x$-direction results in a beam width of $60^\circ$ in the H-plane. Radiation patterns in the E-plane ($x=0$) are characterized in detail in Section~\ref{sec:leaky_wave_farfield}. 

\section{Tapered aperture illumination}

\begin{figure}[htbp]
	\centering
	\includegraphics[width=0.42\textwidth]{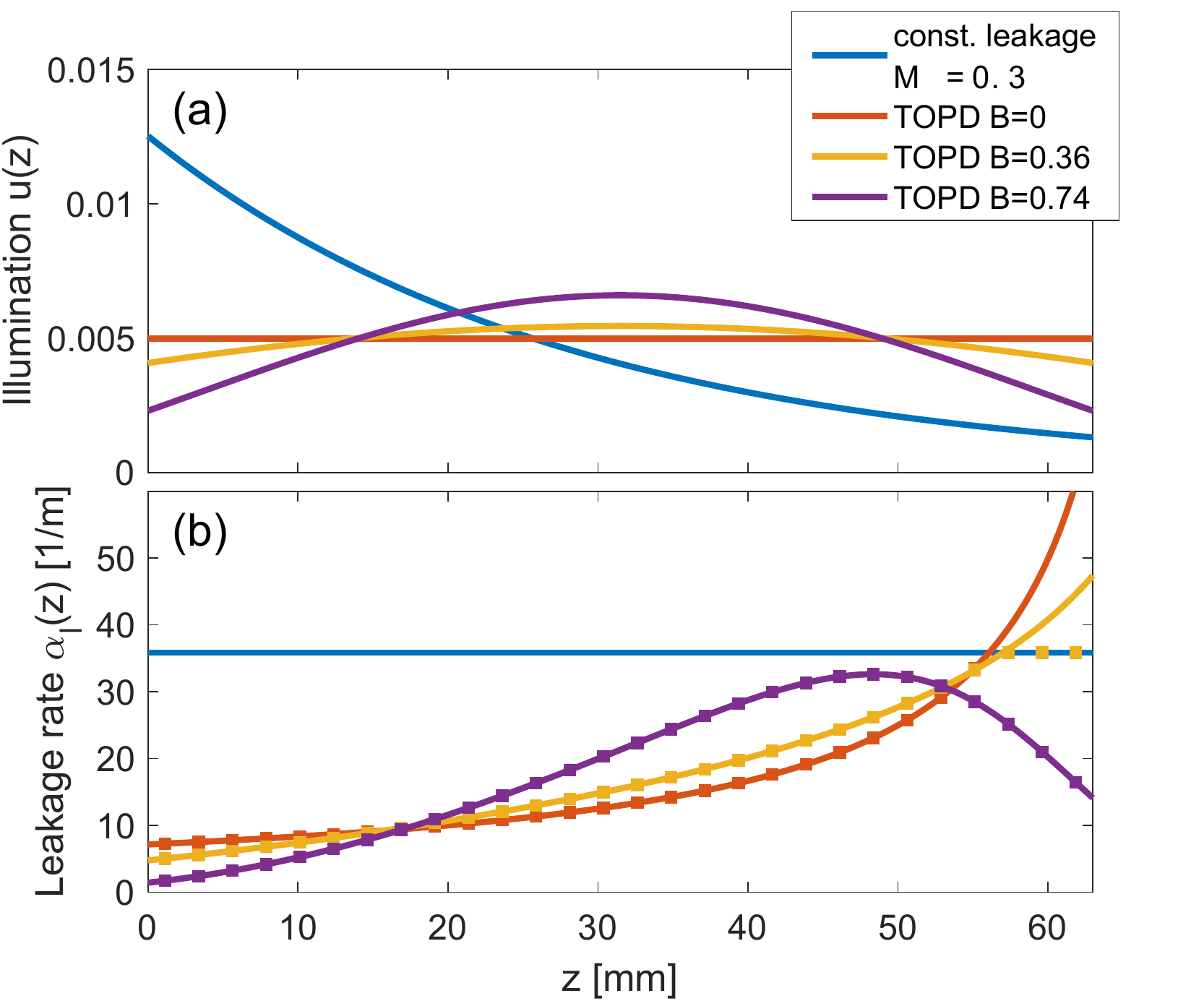}
	\caption{Comparison of constant leakage rate and tapered aperture illumination according to the Taylor one-parameter distribution (TOPD), corresponding to the designed side lobe levels shown in Table \ref{tab:sll}: (a) illumination function $u(z)$ and (b) corresponding leakage rate $\alpha_l(z)$. The square markers indicate the discretization of the continuous leakage rate function, capped at the maximum achievable value.}
	\label{fig:illumination_and_leakage}
\end{figure}

\begin{figure}[htbp]
	\centering
	\includegraphics[width=0.42\textwidth]{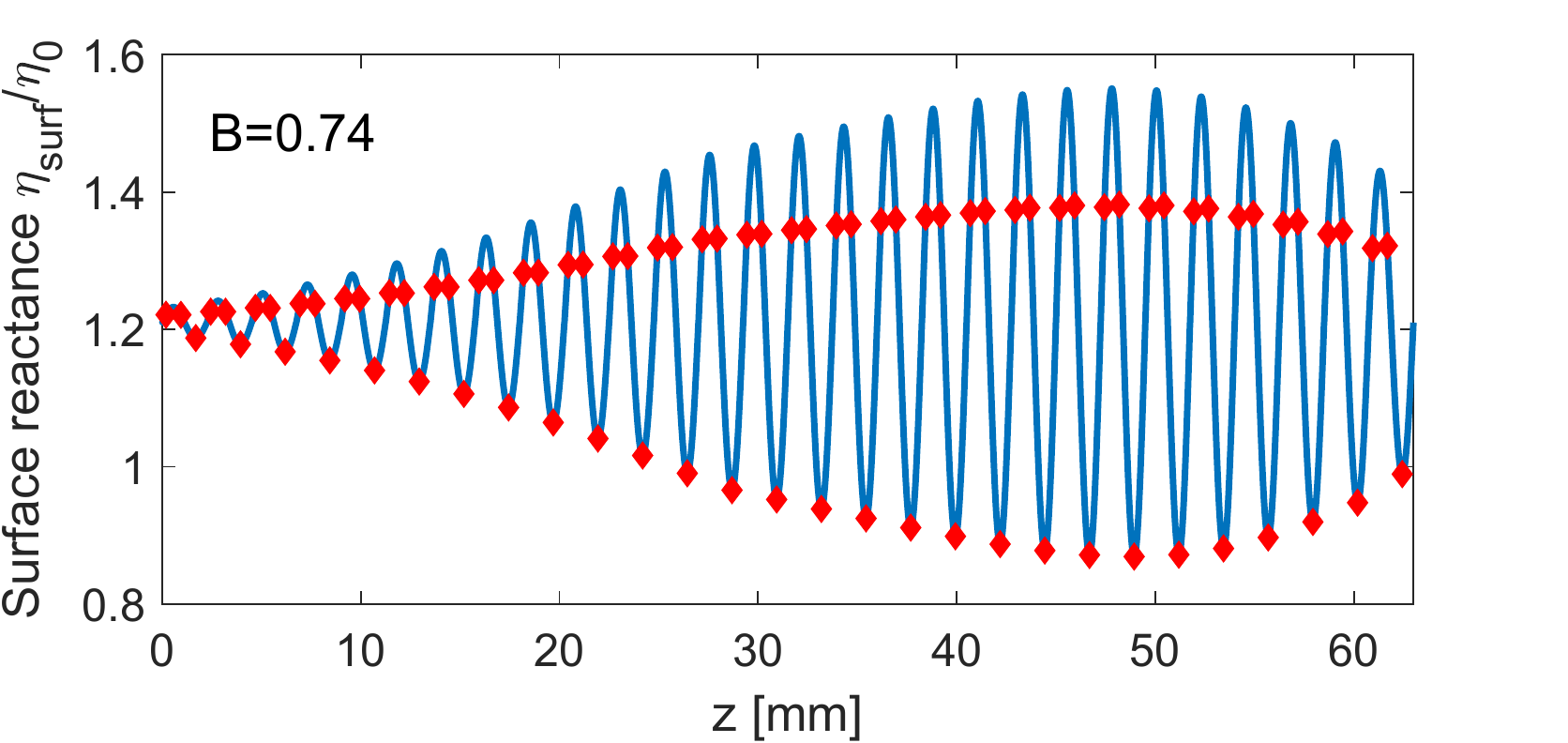}
	\caption{Continuous surface reactance function for the case of strongest tapering $B=0.74$ (blue solid curve) and discretization by 84 unit cells (red square markers). }
	\label{fig:tapered_surface_impedance}
\end{figure}

The array shown in Figure~\ref{fig:prop_const}~(a) represents a leaky wave antenna with constant modulation $M$ and consequently constant leakage factor $\alpha_l$. This results in an aperture illumination distribution $u$ with exponential decay $u(z)\propto \exp(-\alpha_l z)$. Varying the leakage rate along the aperture, the illumination distribution can be controlled effectively. When stipulating an aperture illumination $u(z)$, the required leakage rate $\alpha_l$ can be determined as \cite{jacksonLeakywaveAntennas2008}
\begin{equation}
\label{eq:ap_illu}
    \alpha_l(z)=\frac{\frac{1}{2}|u(z)|^2}{1/\eta_{0} \int_0^L |u(\tau)|^2 d\tau - \int_0^z |u(\tau)|^2 d\tau }.
\end{equation}
Here, $L$ is the aperture length in the $z$ direction. To achieve maximum directivity, a constant aperture illumination can be applied. This will however result in relatively strong side lobes which are 13.26\,dB for an ideal antenna. Using tapered illumination distributions, this side lobe level can be reduced. A common way to control the trade-off between side lobes and main lobe width is to use an aperture illumination according to the Taylor one-parameter distribution \cite{maillouxPhasedArrayAntenna2005,hansenLinearArrays1983}, which has the form
\begin{equation}
    u(z)=\frac{1}{I_0(\pi B)} I_0(\pi B \sqrt{1-(2z/L)^2}),
\end{equation}
where $I_0$ is the modified Bessel function of the first kind of order zero and $B$ is the parameter that controls the side lobe level. In Figure~\ref{fig:illumination_and_leakage}~(a), the exponential illumination obtained with constant leakage rate is compared to the Taylor one-parameter distribution with different values of $B$. For the three cases considered, the edge taper and the expected side lobe level are shown in Table \ref{tab:sll}. Additionally, Figure~\ref{fig:illumination_and_leakage}~(b) shows the corresponding leakage rate calculated with Eq.~\eqref{eq:ap_illu}. This continuous leakage rate function is discretized using 28 supercells of different modulation $M$, which are indicated with square markers. In the case of $B=0$ and $B=0.36$, the required leakage rate exceeds the maximum value that can be implemented with our proposed supercell architecture (Section \ref{fig:supercell_and_band}), therefore we cap the function at the maximum achievable value. The modulated surface reactance function is determined using the dependence of attenuation constant $\alpha$ on modulation $M$ as shown in Figure~\ref{fig:prop_const} and assuming that the attenuation is dominated by leakage, i.e. $\alpha_l \approx \alpha$. It is shown in Figure~\ref{fig:tapered_surface_impedance} for the case of strongest tapering $B=0.74$.

\begin{table}
\begin{center}
\caption{Edge taper and expected side lobe level \cite{maillouxPhasedArrayAntenna2005}}
\label{tab:sll}
\begin{tabular}{ |c | c | c | c|}
\hline
 $B$ & Edge taper [dB]  & Side lobe level [dB] \\ \hline \hline
 0.00 & 0.0 & 13.3 \\  
 0.36 & 2.5 & 15.0 \\ 
 0.74 & 9.2 & 20.0 \\ \hline
\end{tabular}
\end{center}
\end{table}

\section{Experimental demonstration}
\label{sec:LWA_experiments}

\subsection{Sample design and feeding}
\label{sec:contacts}

\begin{figure*}[htbp]
	\centering
	\includegraphics[width=0.95\textwidth]{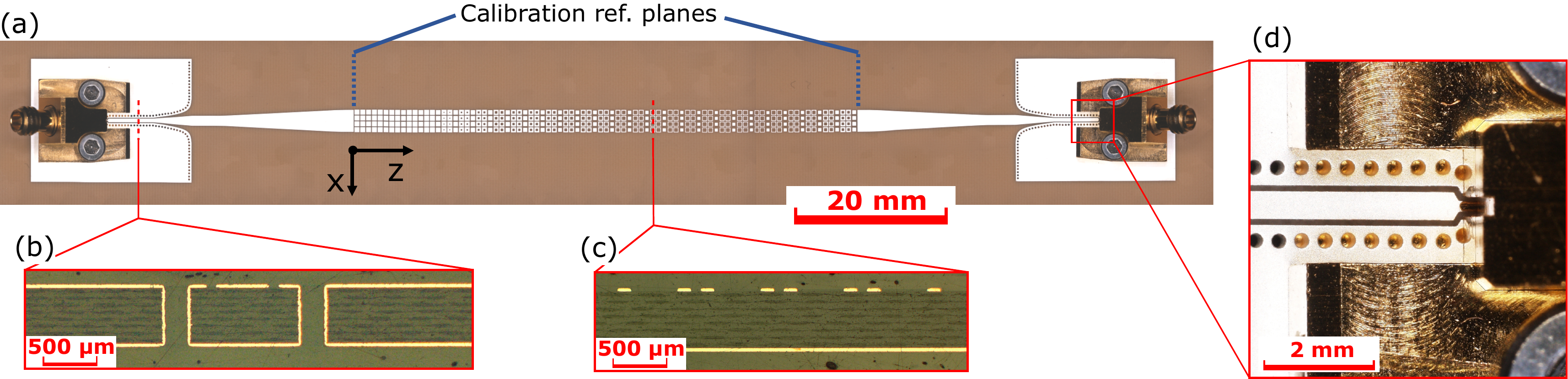}
	\caption{Microscopy pictures showing (a) the top view of the complete tapered leaky-wave antenna ($B=0.74$), (b) cross sectional view in XY plane of a coplanar waveguide section, (c) cross sectional view in XY plane of the patch array and (d) a detailed view of the connector center. During far-field and near-field measurements, the sample was fed from the left port and the right port was connected to a matched load.}
	\label{fig:sampleLWAIII}
\end{figure*}

Five antenna samples were fabricated, among them one surface reactance with zero modulation $M=0$ and one with maximum achievable modulation $M=0.3$. These two samples with constant modulation were fabricated to facilitate the comparison of simulation and measurement. Additionally, three antenna samples with tapered aperture illumination ($B=0$, $B=0.36$ and $B=0.74$) as described in Figure~\ref{fig:illumination_and_leakage} were fabricated. The sample with $B=0.74$ is shown in Figure~\ref{fig:sampleLWAIII}~(a). The samples were fed with commercial solderless coaxial connectors (\textit{Rosenberger 01K80A-40ML5}) which form a transition from 1.0\,mm coaxial cable to grounded coplanar waveguide on the circuit board. With a tapered transition to 50\,$\Omega$ microstripline and a second taper to 30\,$\Omega$ microstripline, this coplanar feed was connected to the patch array. A commercial low loss substrate (\textit{Isola Astra MT77}) with $\epsilon=3.0$ and $\tan \delta=0.0017$ was used, and the conductor pattern was coated with immersion silver. 

Microscopy pictures of polished circuit board cross sections are shown in Figure~\ref{fig:sampleLWAIII}~(b) and (c), illustrating the coplanar waveguide and the patch array, respectively. The via fence of the coplanar waveguide was made with vias of 200\,$\upmu$m diameter. Given that the average surface reactance of the patch arrays is $X'=1.2$, the fundamental mode of a 30\,$\Omega$ microstripline and the TM$_0$ surface wave mode are approximately matched. This is important because the junction of the feed line and modulated reactance surface can lead to significant spurious scattering \cite{olinerLeakywaveAntennas1993}. While mounting the connectors to the board, the proper alignment of the coax center pin was verified using an enhanced focal image microscope. One of these microscope pictures is shown in Figure~\ref{fig:sampleLWAIII}~(d). Using such microscopy images, crucial geometrical dimensions such as the substrate and copper thickness, the microstripline widths and patch sizes were inspected. The corresponding discrepancy between measured and designed dimensions ranges from 2\% to 5\%.

\subsection{S-Parameter measurements and calibration}
\label{sec:s_parameters}

\begin{figure}[htbp]
	\centering
	\includegraphics[width=0.4\textwidth]{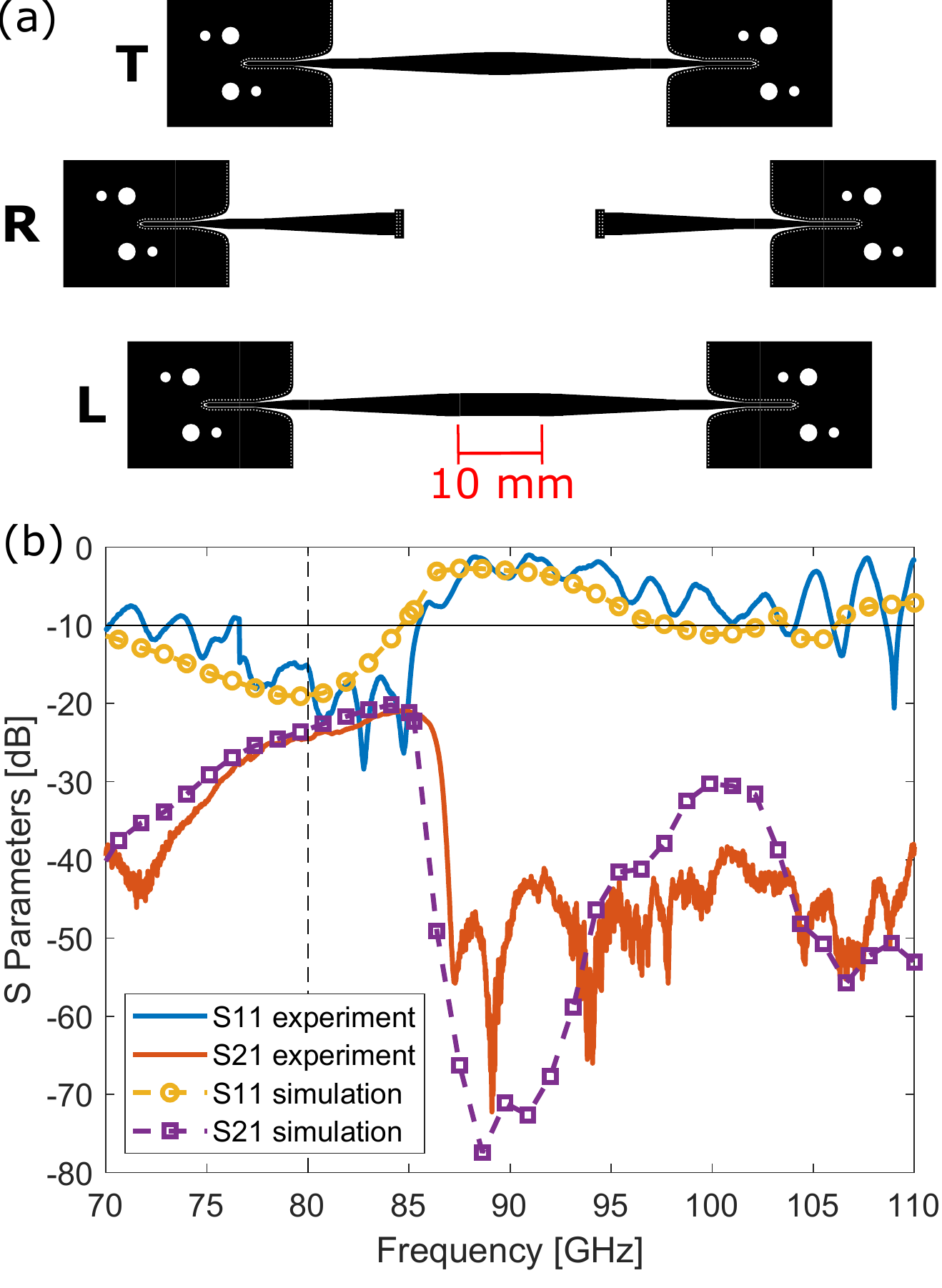}
	\caption{(a) Calibration standards through, reflect (short) and 10\,mm line. The 0.55\,mm and 4.50\,mm lines are not shown. (b) S-parameter from measurement and full-wave simulation for uniformly modulated leaky-wave antenna with $M=0.3$. Due to the multiline TRL calibration, the port reference planes are as shown in Figure \ref{fig:sampleLWAIII} (a). The -10\,dB threshold is marked with a horizontal black solid line and the design frequency with a vertical dashed line.}
	\label{fig:s_par_maxmod}
\end{figure}

The first step of the experimental characterization is the measurement of S-paramters. In order to remove the influence of the connectors, a multiline TRL (through reflect line) calibration \cite{marksMultilineMethodNetwork1991,degrootMultilineTRLRevealed2002} was performed using a custom calibration kit, fabricated on the same PCB panel as the antennas. This calibration brings the reference plane to the end of the 30\,$\Omega$ microstripline as shown in Figure~\ref{fig:sampleLWAIII}. A short was used as reflect standard and the lengths of the lines were chosen as 0.55\,mm, 4.50\,mm and 10\,mm. The through, reflect and one of the line standards are shown schematically in Figure~\ref{fig:s_par_maxmod}~(a).

Results for the sample with maximum modulation $M=0.3$ are shown in Figure~\ref{fig:s_par_maxmod}~(b). With the initial design parameters, the simulated transmission $S_{21}$ was systematically about 2\,dB higher than the measured one. Given that many other geometrical parameters have been verified, we believe that the reason for this discrepancy is that the surface roughness is higher than assumed in the initial design. Therefore, we have adjusted the effective conductivity in the full-wave simulations of this Section from $2\times10^6\,$S/m to $8\times10^5\,$S/m, corresponding to RMS roughness values of 1.5\,$\upmu$m and 2.0\,$\upmu$m, respectively. A similar value was found in previous works where the RMS roughness was measured with profilometer scans \cite{olkHighEfficiencyRefractingMillimeterWave2020}. With the adjusted effective conductivity, the agreement for transmission $S_{21}$ is very good. Significant discrepancies only occur for small values below -30\,dB. At the design frequency of 80\,GHz, most energy is radiated (66\,\% according to simulation) and very little is transmitted to the other port ($S_{21}$ is -24\,dB according to simulation). Above 86\,GHz, $S_{21}$ drops substantially and $S_{11}$ is significantly larger. This marks the beginning of the stopband. Above the stopband, i.e.~above 92\,GHz, the input reflection is significantly larger than at the design frequency. The antennas shown in this work thus operate efficiently using backward waves, not forward waves.

The measured input reflection $S_{11}$ in Fig.~\ref{fig:s_par_maxmod}(b) shows that Fabry-Perot resonances from the coaxial to coplanar transition are not accurately cancelled out by the TRL calibration. This is a common problem when using coaxial connectors in this frequency range \cite{chengMillimeterWaveLowTemperature2014,mondalLeakyWaveAntennaUsing2016,emamiLTCCUltrawidebandPeriodic2018}. Therefore, the accuracy of the measurement of $S_{11}$ is moderate. The overall trend of the input reflection as observed in the full-wave simulation is however confirmed. According to the full-wave simulation, $S_{11}$ is less than -19\,dB at the design frequency 80\,GHz. S-parameters for the tapered leaky-wave antennas 
are qualitatively similar to those of the uniformly modulated antenna, but the effects of the stopband are less pronounced and the Fabry-Perot reflections are slightly stronger.

\subsection{Near-field characterization}

\begin{figure}[htbp]
	\centering
	\includegraphics[width=0.40\textwidth]{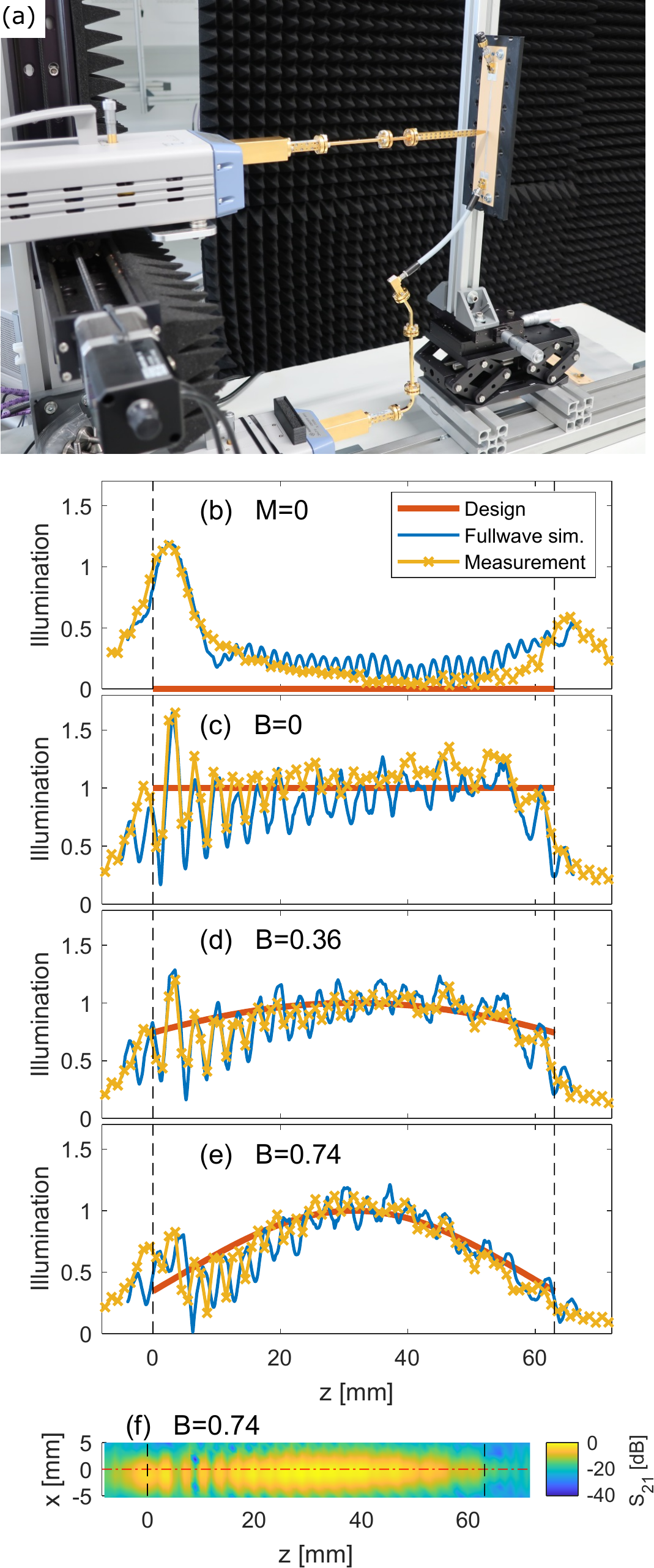}
	\caption{Near-field scan at 80\,GHz. (a) experimental setup, (b) line scan of the unmodulated leaky-wave antenna, (c)-(e) line scan of the tapered leaky-wave antenna with $B=0$, $B=0.36$ and $B=0.74$, respectively and (f) 2D scan of the tapered leaky-wave antenna with $B=0.74$. All scans were taken at a distance of $y=4\,$mm from the aperture.}
	\label{fig:near_field_scan_tapering}
\end{figure}

In order to visualize the aperture illumination and to verify that it has been realized accurately, the antennas were analysed using near-field scanning. The experimental setup is shown in Figure~\ref{fig:near_field_scan_tapering}~(a). The antennas were mounted on a support plate with which a precise alignment to the vertical axis of the scanner was possible. The second port of the antenna was terminated with a 50\,$\Omega$ matched load. More details to the measurement setup can be found in Ref.~\cite{olkHuygensMetasurfaceLens2020}. All scans were taken at a distance of 4\,mm to the antenna aperture and a step width of 1\,mm was used. We compare the amplitude of the electric field $z$-component obtained from simulation and measurement with the illumination function $u(r)$. At this distance, only radiating contributions of the electric field are measured. Therefore, the electric field is expected to be proportional to the illumination $u(r)$.

In Figure~\ref{fig:near_field_scan_tapering}~(b), the result for the unmodulated antenna $M=0$ is shown. Although this antenna is not designed to radiate, a significant field enhancement can be observed around $z=0$ and $z=L$, i.e. at the junction of the antenna feed and the antenna aperture. This field enhancement indicates spurious radiation from the junction. As explained in Section \ref{sec:contacts}, in the design process, this spurious radiation was minimized by choosing suitable dimensions of microstripline, patch array and average surface reactance $X'$. Good agreement between measurement and simulation is discernible. Small variations in the simulated field distribution indicate Fabry-Perot reflections between the two junctions. In the measurement, these variations are smeared out because of the finite size of the near-field probe.

In Figure~\ref{fig:near_field_scan_tapering}~(c)-(e), we show the illumination for the tapered antennas with $B=0$, $B=0.36$ and $B=0.74$. The reflections within the array are stronger, but in all three cases, the overall trend of the electric field amplitude follows the design illumination function. The field enhancement at $z=0$ is smaller for stronger tapering. We note that the electric field in Figure~\ref{fig:near_field_scan_tapering}~(b)-(e) is normalized so that it is on the same scale as the designed illumination function. One normalization constant applies to all measurement curves and a second normalization constant applies to all simulation curves. A 2D scan of the tapered leaky-wave antenna with $B=0.74$ that includes the complete aperture is shown in Figure~\ref{fig:near_field_scan_tapering}~(f). This confirms that the field distribution is relatively uniform across the $x$ dimension, with the expected field decay towards the aperture edges.

\subsection{Far-field characterization}
 \label{sec:leaky_wave_farfield}

The radiated far-fields in the range $\theta=-90...+90^\circ$ of the E-plane were measured using a bistatic facility which has been reported in depth in Ref.~\cite{olkHighlyAccurateFullypolarimetric2017}. The antenna under test and the corresponding frequency extender were placed into the center while the receiving antenna was rotating around it at a distance of 1\,m. As with the near-field characterization, the second port of the antenna was terminated with a 50\,$\Omega$ matched load. The measured gain was determined by normalizing the received power with the power received from a reference horn antenna with 25\,dB of gain. Additionally, the losses from the feeding line and the coaxial cables were characterized separately and subtracted. For this separate characterization, the through from Figure~\ref{fig:s_par_maxmod}~(a) was compared with a waveguide through. Therefore, the reference plane from which the gain is measured is the left calibration reference plane indicated in Figure~\ref{fig:sampleLWAIII}~(a). Due to this normalization, we estimate that the error in the absolute value of the measured gain is about $\pm$1\,dB. Relative errors within one measurement curve are however significantly lower, which means that the side lobe level for instance is determined quite accurately. 

\begin{figure}[htbp]
	\centering
	\includegraphics[width=0.42\textwidth]{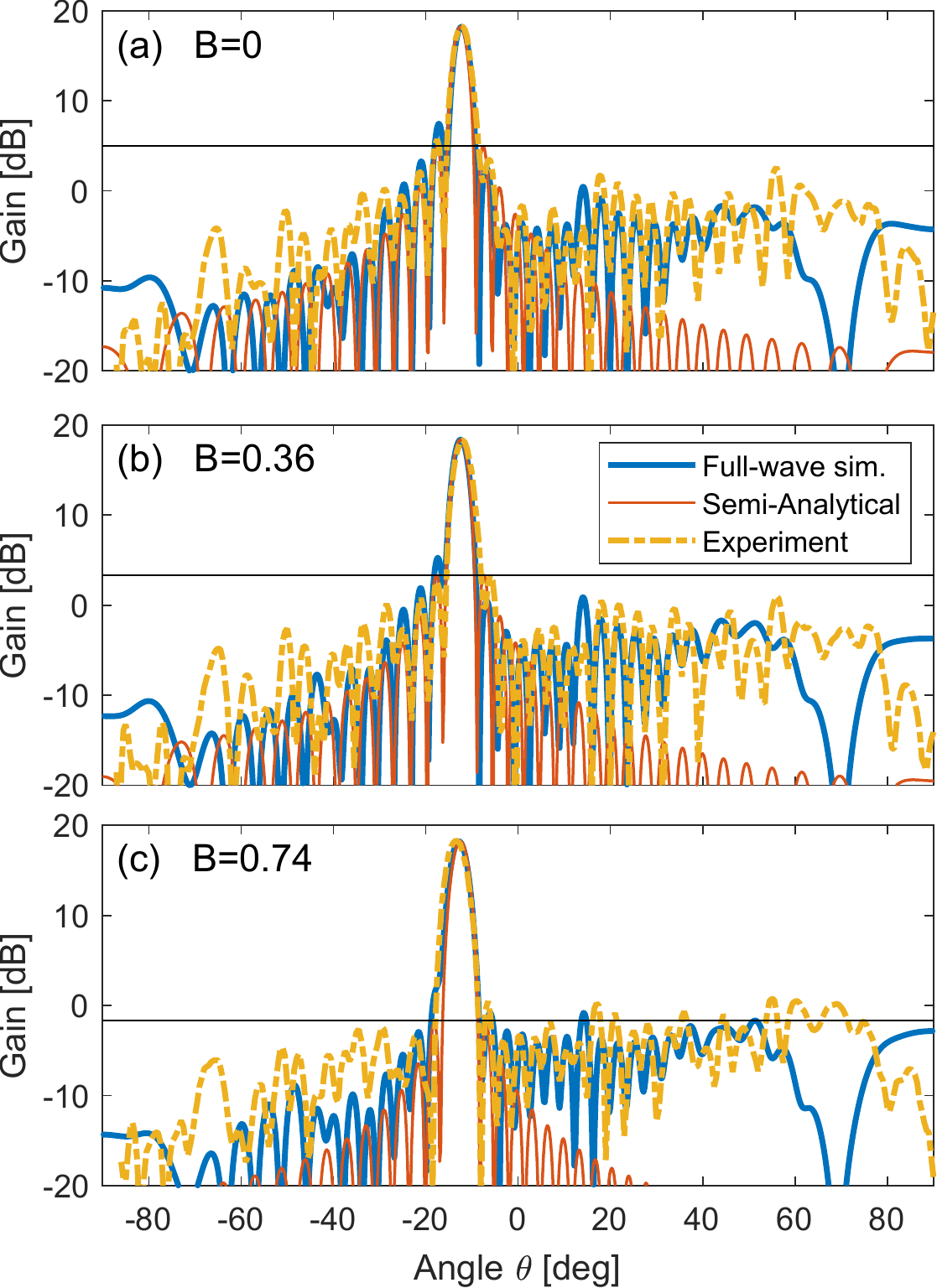}
	\caption{Designed, simulated and measured far-field of the tapered leaky-wave antenna with $B=0$, $B=0.36$ and $B=0.74$, respectively. The design parameter $B$ corresponds to different side lobe levels as shown in Table~\ref{tab:farfield_results} and the expected side lobe level is marked by a horizontal black line. The frequency of operation is 80\,GHz.}
	\label{fig:farfield_tapered_leaky}
\end{figure}

The resulting far-field pattern at the design frequency of 80\,GHz of the tapered leaky-wave antennas is shown in Figure~\ref{fig:farfield_tapered_leaky}. We compare the gain of the antenna obtained with full-wave simulations and from the measurements with the ideal response of the corresponding aperture illumination $u(z)$. The ideal response is calculated as shown in Ref.~\cite{maillouxPhasedArrayAntenna2005} according to 
\begin{equation}
    R(\theta)=\int_0^L u(z) \exp(j k_0 z \sin(\theta) ) dz.
\end{equation}
and it is scaled to match the measured maximum gain. In all three cases, the measured gain reaches its maximum of 18\,dB at about $\theta=12.5^\circ$. The agreement between measured and simulated gain is very good. The side lobe level obtained from the ideal response (Table \ref{tab:sll}) is marked with a horizontal black line. The gain obtained from full-wave simulation and measurement is slightly higher than this for some side lobes. However, the quality of the far-field pattern is very good and the effectiveness of tapering is clearly visible. Within the measured angular range, the cross polarization is less than ${-28}$\,dB compared to the main lobe maximum in all three cases. The side lobe levels and the 3\,dB main lobe widths are shown in Table \ref{tab:farfield_results}. With increasing taper, the main lobe width increases as expected. This result confirms that the design parameter $B$ can control the trade-off between side lobe level and main lobe width.

\begin{table}
\begin{center}
\caption{Far-field results: Side lobe level (SLL) and 3\,dB main lobe width (MLW) in comparison.}
\label{tab:farfield_results}
\begin{tabular}{ |c | c | c | c | c|}
\hline
 \multirow{2}{*}{$B$} & SLL       & SLL Exper-  & MLW Semi-        & MLW Exper- \\
     & Design [dB]     & iment [dB]  & analytical [deg] & iment [deg] \\ \hline \hline
 0.00 & 13.3 & 12.6 & 2.9 & 3.4 \\  
 0.36 & 15.0 & 15.1 & 3.4 & 3.7 \\ 
 0.74 & 20.0 & 17.6 & 3.6 & 4.3 \\ \hline
\end{tabular}
\end{center}
\end{table}

\begin{figure}[htbp]
	\centering
	\includegraphics[width=0.42\textwidth]{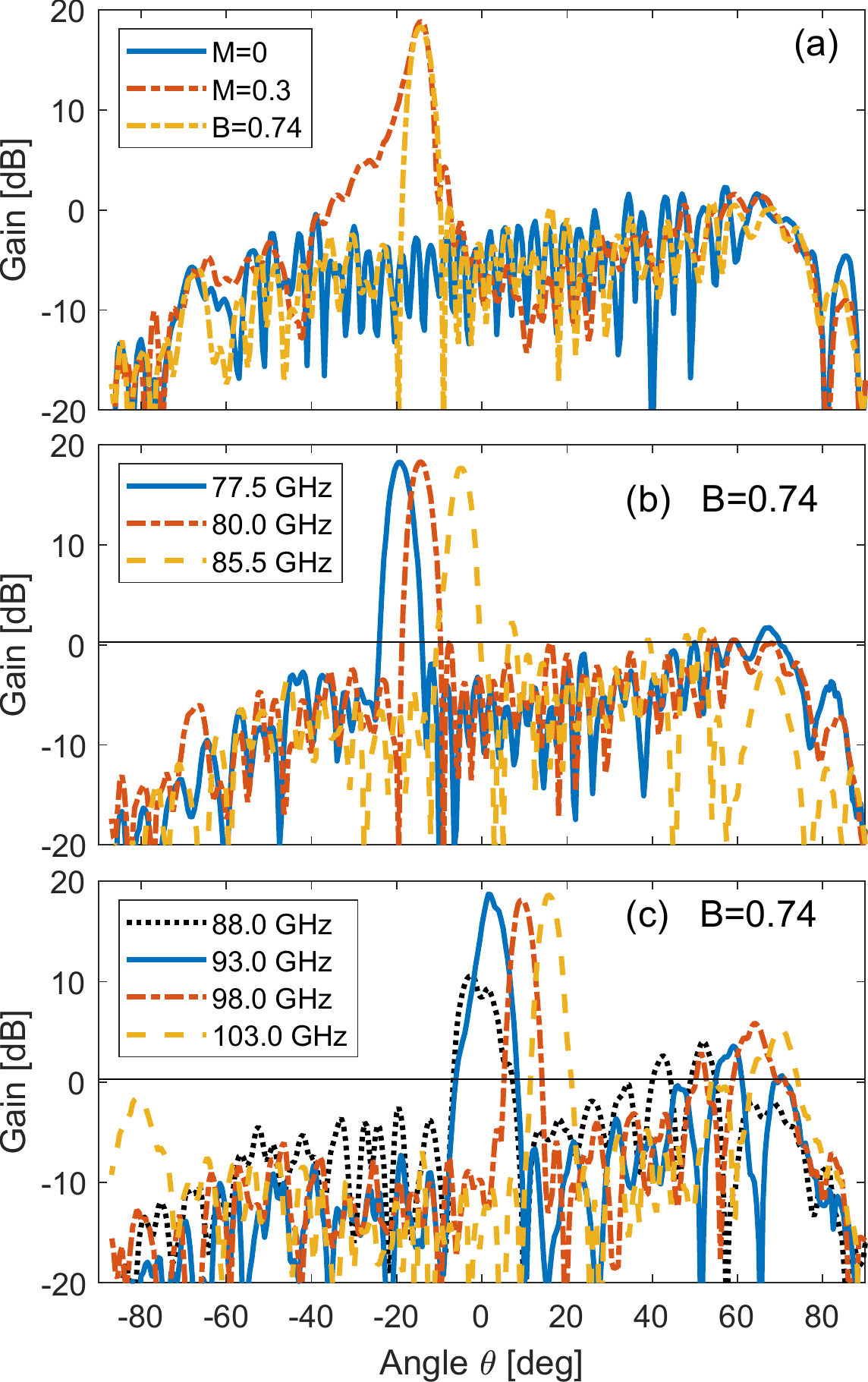}
	\caption{(a) Comparison of the measured gain of the unmodulated reactance surface $M=0$, the reactance surface with maximum achievable modulation $M=0.3$ and the tapered reactance surface with $B=0.74$. (b) Measured gain of the tapered reactance surface with B=0.74 at three different frequencies below the stopband. (c) Measured gain of the same antenna within and above the stopband. The horizontal black line marks the measured side lobe level at the design frequency.}
	\label{fig:lwaIIIthreeFreq}
\end{figure}

Spurious scattering from the junction between the feed and reactance surface has been minimized in the design (Section \ref{sec:coarseSMRS} and \ref{sec:contacts}). In Figure~\ref{fig:lwaIIIthreeFreq}~(a), we show measured far-field patterns of the unmodulated ($M=0$) and modulated  ($M=0.3$) reactance surface and it is apparent that the spurious radiation from this junction contributes mainly to the angular range $+30^\circ$ to $+70^\circ$.  It causes side lobes on the order of 18\,dB and it is visible both in the full-wave simulation and in the measurement. Therefore, using taper larger than $B=0.74$ leads to no further improvement in the side lobe level. The level of spurious scattering from the junction is relatively low, considering that no complex feeding or dedicated surface wave launcher has been used as shown for instance in Refs.~\cite{maCollimatedSurfaceWaveExcitedHighImpedance2017,kuznetcovPrintedLeakyWaveAntenna2019,oraiziDesignWidebandLeakyWave2018}. Additionally, in this figure, the tapered antenna ($B=0.74$) and the antenna with constant modulation ($M=0.3$) are compared.

We further analyse the radiation performance of the  leaky-wave antenna with strongest tapering ($B=0.74$) throughout the  measured frequency range. In Figure~\ref{fig:lwaIIIthreeFreq}~(b), we show the far-field pattern at three different frequencies around the design frequency. This frequency range, from 77.5\,GHz to 85.5\,GHz, is entirely below the stopband, which means that it is in the backward-wave regime. Here, the main lobe is steered from -19.5$^\circ$ to -5.0$^\circ$ and the side lobes of the antenna are below 16\,dB. The horizontal black line marks the measured side lobe level of the antenna at the design frequency, which is only slightly exceeded at other frequencies within this range.

In Figure~\ref{fig:lwaIIIthreeFreq}~(c), we show far-field patterns of the same antenna within and above the stopband. The black dotted line indicates the far-field at 88.0\,GHz, in the middle of the stopband, where the performance of the antenna is significantly degraded. Here, the antenna gain is reduced by 8\,dB. For the frequencies above the stopband which are shown here, the maximum of the main lobe is similar to that around the design frequency, it is only the side lobe suppression that is less effective. This is due to the leakage rate being different at this frequency range compared to the design frequency (Fig.~\ref{fig:prop_const}) which causes the aperture illumination to differ from the design.

\begin{figure}[htbp]
	\centering
	\includegraphics[width=0.42\textwidth]{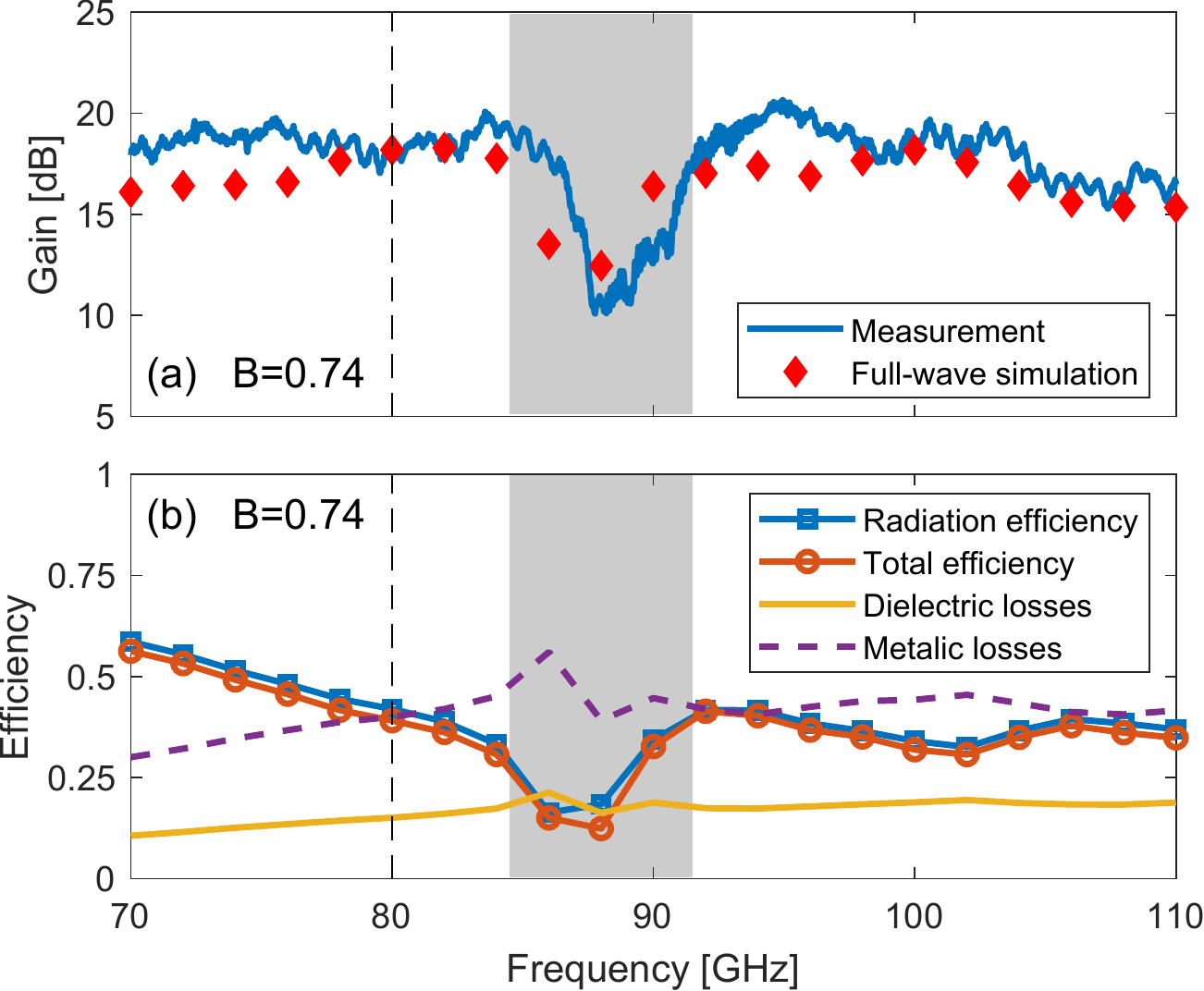}
	\caption{(a) Measured and simulated gain versus frequency and (b) simulated efficiency with dielectric and metallic losses. The grey shaded area indicates the stopband.}
	\label{fig:efficiency_gain}
\end{figure}

In Figure~\ref{fig:efficiency_gain}~(a), we show how the maximum gain of the antenna varies with frequency. This data indicates that the stopband is narrower than estimated with the supercell analysis given in Figure~\ref{fig:supercell_and_band}~(d) where perfect periodicity was assumed. Outside the stopband, the gain is relatively stable. Within the designed frequency range of 77.5 to 85.5\,GHz, the measured gain varies by less than 2\,dB. In Figure~\ref{fig:efficiency_gain}~(b), we show the simulated dissipation losses normalized to the stimulated power and the efficiency of the antenna. The dissipation losses are comparatively high, especially those caused by the rough surface of metallic elements. This causes the efficiency to be less than 45\,\% within the designed frequency range, compared to 66\,\% in the case of maximum modulation (Section~\ref{sec:s_parameters}).

\subsection{Comparison with literature}

\begin{table}[]
\begin{center}
\caption{Antenna with $B=0.74$ in comparison with leaky-wave antennas reported previously.}
\label{tab:performance_comparison}
\begin{tabular}{|c|c|c|c|c|c|}
\hline
\multirow{2}{*}{Reference}                                     & Frequency &  Gain  & Efficiency & SLL  & \multirow{2}{*}{Fabrication}                \\ 
& {[}GHz{]} & {[}dB{]} & [\%] & {[}dB{]} & \\
\hline \hline
\cite{patelPrintedLeakyWaveAntenna2011} & 9-11               & 18                 & n.a.  & 10           & PCB                       \\ \hline 
\cite{kuznetcovPrintedLeakyWaveAntenna2019} & 23-24               & 11                 & 85  & 10           & PCB                       \\ \hline 
\cite{yangTaperedUnitCell2018} & 9.5-10.5               & 17                 & 30 & 14           & PCB                       \\ \hline \hline
\multirow{2}{*}{\cite{baiSinusoidallyModulatedLeakyWave2016}} & \multirow{2}{*}{50-85}               & \multirow{2}{*}{14.2}               & \multirow{2}{*}{60-75}  & \multirow{2}{*}{10}           & Inset dielect.  \\ 
& & & & & waveguide \\ \hline 
\cite{chengMillimeterWaveLowTemperature2014} & 58-67               & 12                 & 60-80  & 10           & LTCC                       \\ \hline 
\cite{chengMillimeterWaveLowTemperature2014} & 90-98               & 22                 & 50-65   & 13           & LTCC                       \\ \hline 
\multirow{2}{*}{\cite{mondalLeakyWaveAntennaUsing2016}}       & \multirow{2}{*}{96-108}              & \multirow{2}{*}{12.7}               & \multirow{2}{*}{n.a.}       & \multirow{2}{*}{11}         & Perforated       \\ 
& & & & & dielectric \\ \hline
\multirow{2}{*}{\cite{sanchez-escuderosPeriodicLeakyWaveAntenna2013}} & \multirow{2}{*}{38-42} & \multirow{2}{*}{16} & \multirow{2}{*}{71} & \multirow{2}{*}{16} & PCB with \\ 
& & & & & reflector \\ \hline \hline
This work & 77.5-85.5 & 18 & 33-45  & 18 & PCB \\ \hline

\end{tabular}
\end{center}
\end{table}

In Table~\ref{tab:performance_comparison}, we compare the performance of the antenna with the strongest tapering ($B=0.74$) with related experimentally demonstrated leaky-wave antennas presented in the literature. This table includes the designed frequency range, the maximum gain, the efficiency and the measured side lobe level (SLL) at the design frequency. In Refs.~\cite{patelPrintedLeakyWaveAntenna2011}, \cite{kuznetcovPrintedLeakyWaveAntenna2019} and \cite{yangTaperedUnitCell2018}, leaky-wave antennas with a similar architecture fabricated on PCBs were reported. These antennas operate at significantly lower frequencies and they feature side lobe levels of 10\,dB (Ref. \cite{patelPrintedLeakyWaveAntenna2011} and \cite{kuznetcovPrintedLeakyWaveAntenna2019}) and 14\,dB in the case of Ref.~\cite{yangTaperedUnitCell2018}, where amplitude tapering was applied.

Additionally, in Ref. \cite{baiSinusoidallyModulatedLeakyWave2016}, \cite{chengMillimeterWaveLowTemperature2014} and \cite{mondalLeakyWaveAntennaUsing2016}, leaky-wave antennas operating at mm-wave frequencies were reported. As mentioned in Section~\ref{sec:LWAintroduction}, in order to overcomes the high losses, and fabrication constraints of PCB fabrication, the authors used fabrication methods which are comparatively costly. %
The Goubau-type leaky-wave antenna from Ref. \cite{sanchez-escuderosPeriodicLeakyWaveAntenna2013} operates around 40\,GHz, features high efficiency and was manufactured on a PCB. To achieve directive radiation, the authors placed the antenna on a metallic reflective plate with a 5\,mm foam spacer. This overview table confirms the aforementioned benefits of our proposed approach. It facilitates low-profile designs and planar fabrication of leaky-wave antennas operating at mm-wave frequencies while enabling effective aperture control and side lobe suppression. These benefits come at the expense of a relatively low antenna efficiency. 
We note that tapered leaky-wave antennas with large apertures often have low efficiencies due to the extended propagation distance of the surface wave along the aperture. The same issue was observed for instance in Ref.~\cite{yangTaperedUnitCell2018}. The efficiency of the proposed antenna could potentially be further optimized by improving the quality of the conductor surface or by reducing the amount of energy that is absorbed in the matched load which is connected to the second port (here 10\,\% at 80\,GHz). 

\section{Conclusion}

In this paper, 1D leaky-wave antennas operating at W-band frequencies are designed, fabricated and experimentally demonstrated. By using a unit cell architecture with very coarse discretization we circumvent previous difficulties for higher frequency designs such as limited reactance ranges. In particular, the minimum feature size of printed circuit board processes can be taken into account while maintaining the high flexibility of aperture control of sinusoidally modulated reactance surfaces. We demonstrate the versatility of this concept by presenting three different antennas with different levels of aperture tapering. By means of near-field scanning, the designed aperture illumination is verified. These scans also show that the most significant part of spurious radiation originates at the junction of the antenna feed and reactance surface. Far-field measurements confirm that this spurious radiation as well as side lobes are effectively suppressed to a level of ${-18}$\,dB of the main lobe power.


%


\section*{Acknowledgment}

This work was financially supported by the Australian Research Council (Linkage Project LP160100253), the  Luxembourg Ministry of the Economy (grant CVN 18/17/RED) and the University of New South Wales (UIPA scholarship). A.~E.~O.~thanks Cédric Amorosi for the support in microscopy analysis and Christian Pauly for valuable discussions.


\appendices
\section{} \label{sec:appendix_1}

\begin{figure}[htbp]
	\centering
	\includegraphics[width=0.42\textwidth]{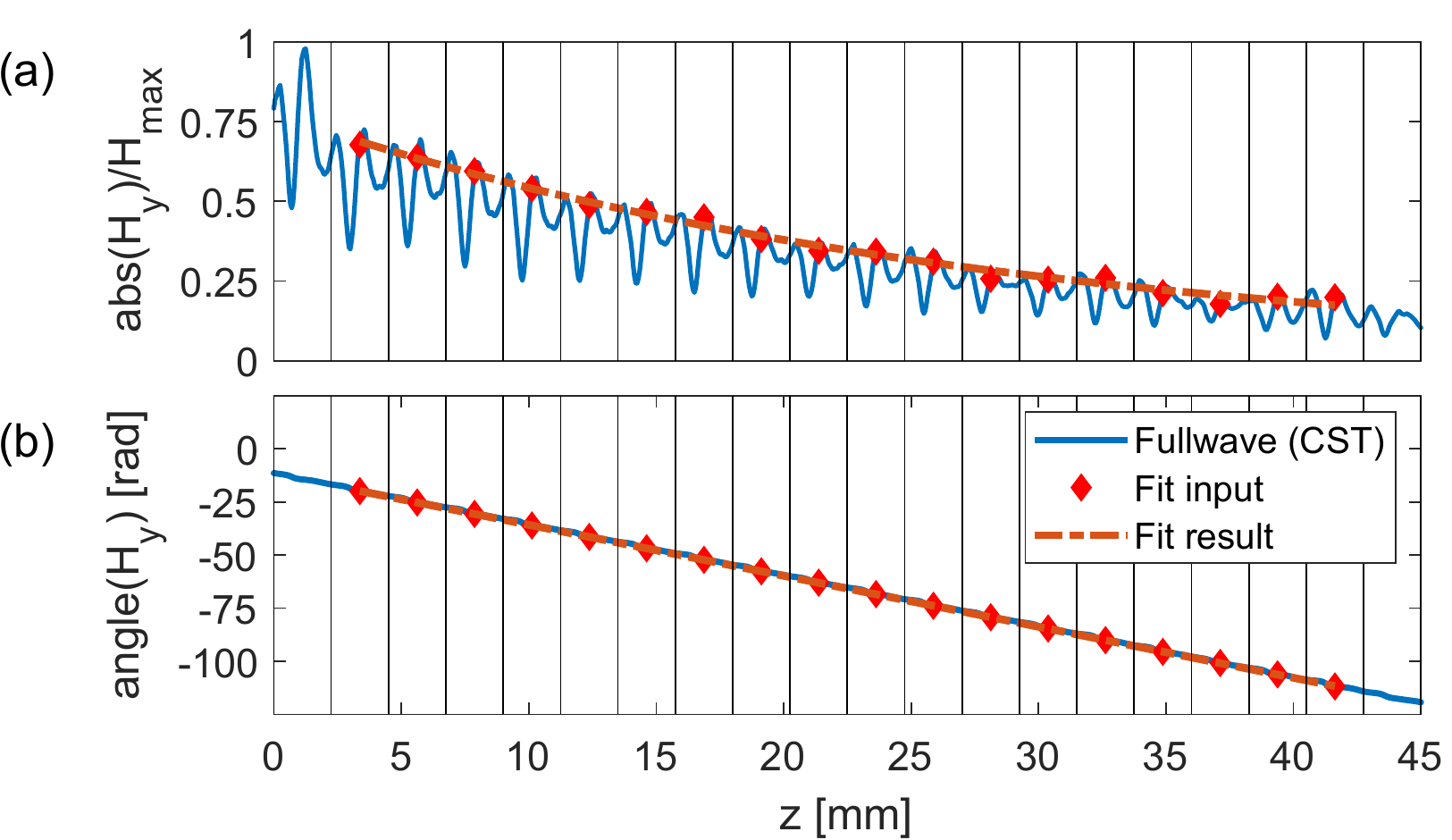}
	\caption{Evaluation of (a) propagation constant $\beta$ and (b) leakage rate $\alpha$. The data shown in this Figure corresponds to a constant modulation depth of $M=0.3$ and a Frequency of operation of 80\,GHz, and it was obtained through full-wave simulation.}
	\label{figx_fit}
\end{figure}

In Figure~\ref{figx_fit}, we show a fit to the H-field with which the propagation constant $\beta$ and the leakage rate $\alpha$ was determined. In particular, the blue lines correspond to the transverse component of the magnetic field $H_y$ determined in full-wave simulations. The red markers correspond to a value of this magnetic field evaluated at the center of every supercell which served as input for the fit.

\ifCLASSOPTIONcaptionsoff
  \newpage
\fi



%
%
%


\bibliographystyle{IEEEtran}
\bibliography{LWA}

%


\begin{IEEEbiography}[{\includegraphics[width=1in,height=1.25in,clip,keepaspectratio]{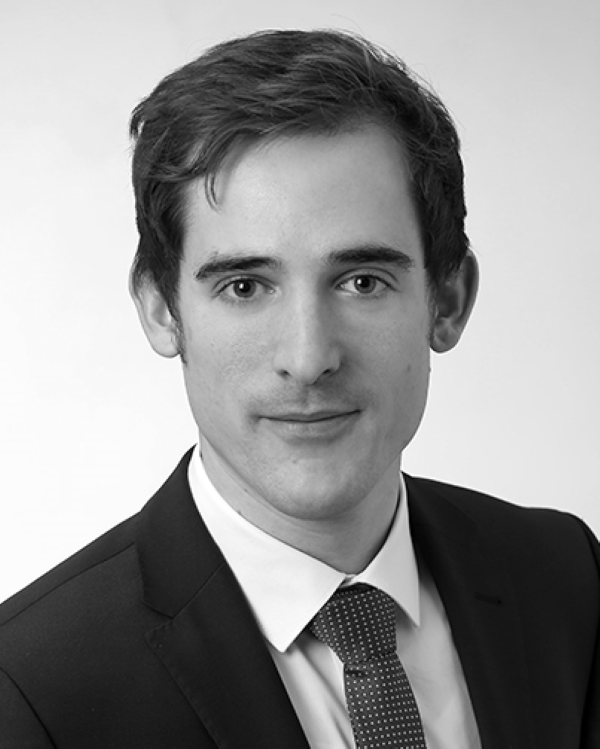}}]{Andreas E. Olk}
	 received the B.Sc and the M.Sc. degree in physics from RWTH University, Aachen, Germany, in 2012 and 2014, respectively. Since 2015, he is affiliated with IEE S.A., Luxembourg. Since 2018, he is pursuing a Ph.D. in electrical engineering with the University of New South Wales, Canberra, Australia.  His current research interests include metamaterials, millimeter-wave technology, radar and automotive sensing.
\end{IEEEbiography}

\begin{IEEEbiography}[{\includegraphics[width=1in,height=1.25in,clip,keepaspectratio]{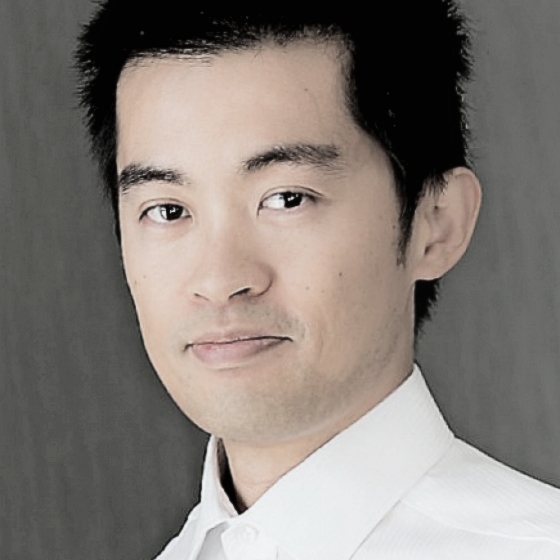}}]{Mingkai Liu}
	 received the B.Sc in optical information (2008) and the M.Sc. in optics (2011) from Sun Yat-sen University, China, and then the Ph.D in physics (2015) from the Australian National University, Australia. He was a research fellow at the Nonlinear Physics Center of the Australian National University from 2015 to 2019 and recently joined industry as a data scientist. His research interest is to develop time-varying, reconfigurable, and ultra-compact electromagnetic devices, by combining physics and machine-learnings.
\end{IEEEbiography}

\begin{IEEEbiography}[{\includegraphics[width=1in,height=1.25in,clip,keepaspectratio]{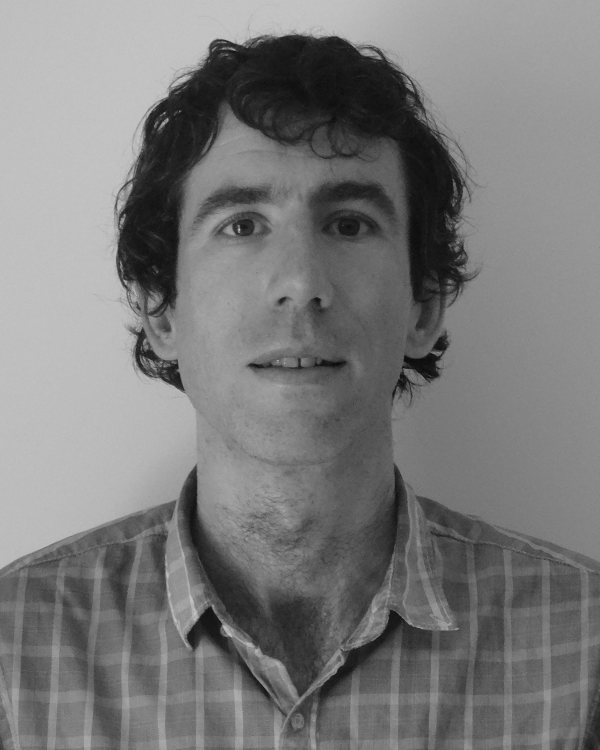}}]{David A. Powell} (M’05-SM'17) is a Senior Lecturer at the School of Engineering and Information Technology, University of New South Wales, Canberra, Australia.
    His research on metamaterials has covered the microwave, millimeter-wave, terahertz, and near-infrared wavelength ranges, in addition to work on acoustic metamaterials.	He received the Bachelor of Computer Science and Engineering from Monash University, Melbourne, Australia, in 2001, and the Ph.D. degree in Electronic and Communications Engineering from RMIT University, Melbourne, Australia in 2006. Between 2006 and 2017, he was a researcher with the Nonlinear Physics Centre at the Australian National University. 
\end{IEEEbiography}




\end{document}